\documentclass[%
 reprint,
 amsmath,amssymb,
 aps,
prl,
]{revtex4-1}

\usepackage{graphicx}
\usepackage{epstopdf}
\usepackage{dcolumn}
\usepackage{bm}
\usepackage[normalem]{ulem}  

 \usepackage[usenames,dvipsnames]{color}

\begin{document}

\preprint{APS/123-QED}

\title{Classical trajectories in polar-asymmetric laser fields:\\
Synchronous THz and XUV emission
}

\author{Aram Gragossian,$^{1}$ Denis V. Seletskiy,$^{1,2}$ and Mansoor Sheik-Bahae,$^{1}$}

\address{
$^1$University of New Mexico, Physics and Astronomy Dept.,  1919 Lomas Blvd. NE, Albuquerque, NM 87131, USA\\
$^2$Department of Physics and Center for Applied Photonics, University of Konstanz, 78457 Konstanz, Germany\\
}

\begin{abstract}
Synchronous extreme ultraviolet (XUV) and single-cycle terahertz (THz) bursts
are generated in argon plasma induced by intense two-color femtosecond laser
pulses. Correlations between the intensity of the even and odd high-harmonics
and the THz radiation are found by studying the phase-delay between the
excitation pulses at 800 and 400 nm as well as the degree of polar asymmetry in
the incident electric field. Experiments in both the weak- and strong-
polar asymmetric regimes show remarkable agreement with a simple analytical
model based on classical electron trajectories in an arbitrary
synthetic electric field.
\end{abstract}

\maketitle


Generation and manipulation of coherent radiation at extreme wavelengths has
advanced immensely in the recent years. Significant progress has
been made in the field of high harmonic generation (HHG) where the interaction
of intense near-infrared laser pulses with rare-gases produces bursts of
radiation with spectral content extending into the extreme ultraviolet (XUV)
\cite{McPherson1987,Antoine1996} and  soft x-ray
\cite{Drescher2001,Popmintchev2012} regions. Along with the advent of novel
diagnostic schemes \cite{Krausz2009,Chini2014,Taec2014}, this has led to the
generation of isolated sub-100 attosecond pulses
\cite{Goulielmakis2008,Zhao2012} and can potentially support zeptosecond pulses
in the near future \cite{Popmintchev2012}. On the other end of the spectrum,
advances have been made in the generation and detection of intense coherent
sub-harmonic optical waveforms, covering the entire terahertz (THz) frequency spectrum and
extending into the mid- and near-infrared
\cite{Wu1996,Dai2006,Sell2008,
Matsubara2012}. Electromagnetic pulses in both
spectral extremes can be efficiently generated and manipulated using a two-color
excitation scheme where the fundamental driving pulse is combined with its
second harmonic (SH) in a rare gas plasma
\cite{Kim2005a,Mauritson2006,Roskos2007,Brugnera2011,Zhao2012,Cook2000,Kim2007}.

The generation of high harmonics in rare gases follows a non-perturbative
mechanism, with its salient features captured by a classical three-step model
\cite{Corkum1993,Brabec2000}. In this picture, electrons are liberated in a
tunnel ionization process and accelerated every half cycle of the driving laser
pulse. Depending on their birth time, a fraction of these electrons can acquire
sufficient kinetic energy to trigger high-energy photon emission
upon recombination with the parent ion. This simple, insightful model was
confirmed by a quantum-mechanical approach in the strong-field-approximation
\cite{Lewenstein1994} followed by a detailed conversion efficiency analysis
\cite{Falcao-Filho2009,Falcao-Filho2010}. A rigorous description of the
ionization and the ensuing electronic wavepacket dynamics was performed by
numerical evaluation of time-dependent Schr{\"o}dinger equation (TDSE) \cite{Krause1992,Figueira1997,Bauer2006}.
The polar symmetry of the laser field and the centro-symmetry of
the gaseous media cause the generated bursts of XUV emission to contain only odd
harmonics of the carrier frequency $\omega_0$. This symmetry can be broken by
injecting a small fraction of phase-locked SH field at
2$\omega_0$, resulting in emission of odd as well as even harmonics
\cite{Kim2005a,Dudovich2006,Mauritson2006,Doumy2009,Oishi2006,Brugnera2011}. The
resulting polar asymmetry can be coherently controlled by the relative phase
difference between these fields. Using this scheme, it was shown that the
birth of attosecond pulses can be controlled with high precision
\cite{Dudovich2006,He2010,Dahl2011}.
Two-color excitation has also been employed to generate controlled
bright single-cycle THz radiation in gaseous targets \cite{Cook2000}. While
initial observations were qualitatively consistent with a $\chi^{(3)}$ model
\cite{Cook2000,Kress2004,Xie2006}, a satisfactory quantitative description was
given by a classical macroscopic plasma current model that essentially uses the
first two steps of the 3-step model without recombination
\cite{Kim2007,Kim2008}. In this picture, some of the electrons that do not
return to the parent ion may create a transient net current (following the
pulse envelope) with a bandwidth extending to tens of THz  \cite{Kim2007}. The
asymmetric (drift) trajectories of the ionized electrons adequately describe the
key features of the THz experiments. The TDSE has also been used to analyze
certain experimental features in the THz radiation
\cite{Wang2008,Kim2005a,Dai2009b}.
The intriguing similarities in the physics of XUV and THz generation together
with the practical implications of temporal synchronization provide
motivation for probing their correlations and coherent control. In this
letter, we investigate simultaneous generation of XUV and THz radiation under
conditions of weak and strong polar asymmetry in co-polarized two-color excitation. We
present a simple and intuitive analytical expression based on classical
electron trajectories and tunnel ionization (extended 3-step model) that
well describes the main experimental features of HHG in both
polar asymmetry regimes. This model successfully explains the observed
correlations of XUV emission with the synchronously-generated THz radiation.

\begin{figure}[htb]
  \centerline{\includegraphics[trim=0cm 1cm 0cm 2cm,width=3.2in]{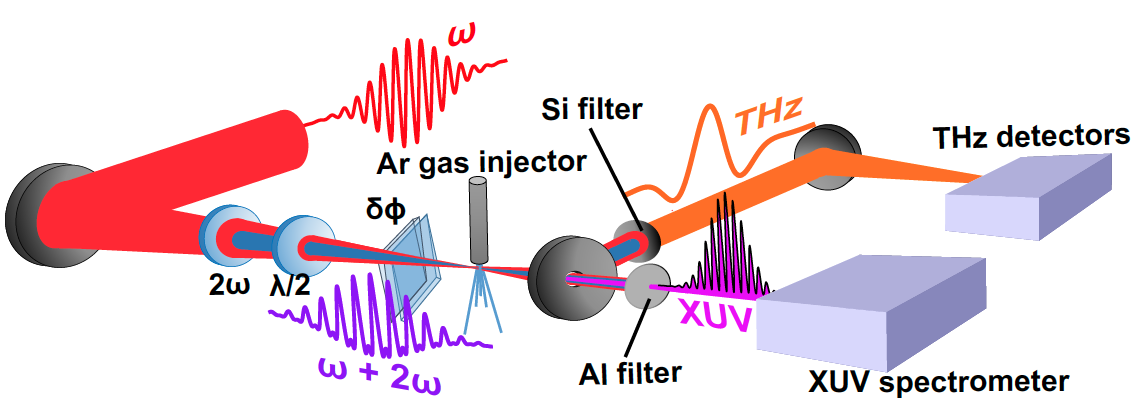}}
  \caption{Schematic of the experimental setup.}
  \label{Fig:setup}
  \end{figure}

The experimental setup for synchronous generation and detection
of XUV and THz radiation is shown schematically in Fig. \ref{Fig:setup}. A train
of 1 kHz, 40 fs pulses with a center wavelength of 800 nm produces high
harmonics when focused onto the output of an argon gas injection nozzle
with peak intensity of $5\times10^{14}$  $W/cm^2$. Good spatial and
temporal coherence properties of the high harmonics are ensured by placing the
nozzle after the focus \cite{Salieres1995}.

The SH of $\omega_0$ is generated in a 150 $\mu m$ type-I BBO
crystal, the orientation and position of which control the conversion efficiency
of SH in the range of 0.5\% to approximately 10\%. Parallel
polarizations of the two fields are ensured by an ultra-thin, true zero-order
half-wave plate for the fundamental wavelength. As expected from symmetry
arguments and confirmed by measurements, this results in linearly polarized
THz and XUV emission. A pair of glass wedges controls the relative phase
$\delta\phi$ and hence the degree of asymmetry between two pulses with an
accuracy $\approx\pi/5$ that is independently calibrated in a separate
experiment \cite{Chudinov1991}. All of these optical components are placed in
a low background pressure chamber to ensure minimal phase-slip between the
fundamental and SH pulses. An off-axis parabolic mirror is located after the gas
nozzle and serves to collimate the THz field out of the vacuum chamber through a
silicon viewport. This window also acts as
an optical filter to remove the excitation light. A 500 $\mu m$ diameter
central hole of the parabolic mirror transmits the
low-divergence XUV beam. It then passes through a 200~nm thick aluminum
filter and is routed to a grazing-incidence XUV spectrometer (McPherson
Inc.) as shown in Fig. \ref{Fig:setup}. Terahertz emission is
field-resolved with electro-optic sampling and direct detected by a
pyroelectric detector. Polarization states of the XUV and THz pulses are ensured
using polarization-dependent anisotropy of the grating reflectivity and a
broadband polarizer, respectively, both with contrast better than 10 to
1.

\begin{figure}[htb]
\centerline{\includegraphics[trim=1cm 1cm 1cm 0.5cm,width=3.1in]{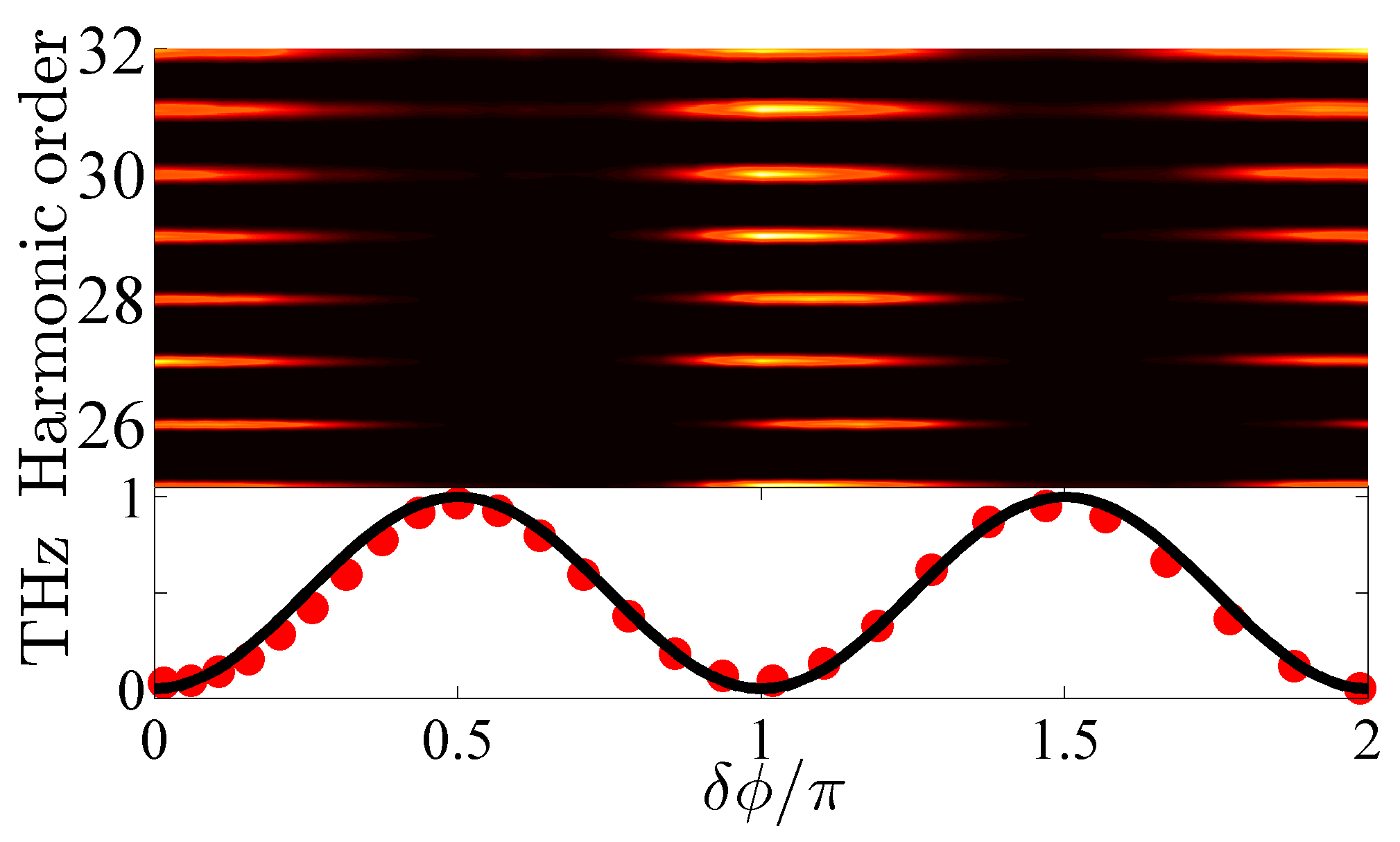}}
\caption{High harmonics (top) and THz energy (bottom) as a function of the relative phase $\delta\phi$. Dots are measurements and the black line is calculated THz using plasma current model.}
\label{Fig:XUVTHz}
\end{figure}

Simultaneous emission of XUV and THz radiation is measured as a
function of phase delay $\delta\phi$ within the synthetic excitation pulse. We
consider two regimes of weak and strong polar asymmetry corresponding to low
($\leq$ 1\% ) and high ($\geq$ 10\%) SH injection, respectively, as
referenced to the average power of the fundamental pulse train. Under the
condition of strong SH injection, the measured HHG spectrum contains even
and odd harmonics with both exhibiting maxima corresponding to
$\delta\phi\approx m\pi$ for all of the harmonic orders (Fig. \ref{Fig:XUVTHz}).
Phase delay also leads to modulation of the THz power, but with the maxima
around $\delta\phi\approx (2m+1)\pi/2$, corresponding to fully polar symmetric
two-color excitation. The anti-correlated nature of the emissions can be
qualitatively understood by considering classical electron trajectories.
According to the 3-step model, those electrons responsible for HHG follow closed
trajectories that
terminate in a collision with a parent ion within a half-cycle of the driving
pulse. This cannot result in any net charge displacement (or current)
at low frequencies.\
\\
THz generation, on the other hand, is described by a classical macroscopic
plasma current model where tunnel ionization is followed by electron
acceleration. This builds with each consecutive half-cycle of the
2-color pulse to generate a net low-frequency
current \cite{Kim2007,Kim2008}. Such accumulation is only possible if the
instantaneous drift velocity ($v$) profile has polar asymmetry. Because
$v(t)\propto\int Edt'$, this corresponds to the case of polar-symmetric
excitation field E with $\delta\phi=(2m+1)\pi/2$ \cite{Kim2007}. The
THz emission is shown to follow $\sin^2(\delta\phi)$, in close agreement with
our observation (Fig. \ref{Fig:XUVTHz}).
It is instructive to develop an analytical  model and test its sensitivity to
other experimental parameters. Corkum's
Simple Man's Model \cite{Corkum1993} is broadened to obtain a compact but
qualitative expression for the HHG spectrum under arbitrary linearly polarized
synthetic excitation. Our
\textquotedblleft extended three-step \textquotedblright (ETS)
model captures the observed details for both
weak and strong asymmetry.

We emphasize that the following \textquotedblleft
intuitive\textquotedblright  treatment of HHG is based on classical
trajectories. By its very nature, it cannot be rigorous as it makes a quantum
leap (literally) by assuming that each re-collision produces a photon in a
deterministic way. Our primary objective is to elucidate the critical role of
the polar asymmetry of the excitation field, i.e.~ understanding the effect of
SH power ($\eta$) and its relative phase delay ($\delta\phi$) with with respect to the
fundamental. We are not concerned with the absolute magnitudes,
conversion efficiency, or propagation evolution including the phase-matching
requirements. We begin with two-color laser pulses having a combined local
(fixed position along the propagation axis) electric field $E(t)=E_1(t)+E_2(t)$
with $E_1(t)=A_0(t)cos(\omega_0t)$ and $E_2(t)=\sqrt{\eta}
A_0(t)cos(2\omega_0t+\delta\phi)$, where $A_0$ is the pulse envelope. After an
ionizing event, the transverse electron trajectories $x(t_i,t_r,\delta\phi)$
are
obtained from solution of the classical equation of motion ($m_0\ddot{x}=-eE$)
given a birth time $t_i$ and no initial momentum. As in the one-color
excitation, the return times ($t_r$) for each trajectory are the solution of
$x(t_i,t_r,\delta\phi)=0$, with a return kinetic energy
$U(t_i,t_r,\delta\phi)=m_0\dot{x}(t_i,t_r,\delta\phi)^2/2=U_pF(t_i,t_r,
\delta\phi)$, where $U_p={e^2A_0^2}/4m_0\omega_0^2$ is the pondermotive energy.
Exploiting the well-known relationship between the field amplitude and the
photon number, we take the spectral amplitude of the emitted harmonics to vary
as $|E_H(\omega)|\propto\sqrt{\hbar\omega\dot{\rho}}$ where $\dot{\rho}(t_i)$
is the ionization rate at a given birth time $t_i$ for electron trajectories
having return energy $U$ that satisfies $U(t_i,t_r,\delta\phi)+I_p=\hbar\omega$
with $I_p$ denoting the ionization potential of the gas. Next, we assign a
relative spectral phase $\omega t_r$ to such a trajectory; this signifies the
distinct re-collision (arrival) times of each 
trajectory \textquotedblleft bunch\textquotedblright in the time domain. The
total local HHG electric field $E_H(\omega,\delta\phi_r)$ is then obtained by
summing the contributions  from all trajectories during the excitation pulse (up
to four per optical cycle of the fundamental) having the same return energy
($\hbar\omega$):

\begin{multline}
E_H\left( \omega,\delta\phi\right) \propto \\
\sum_{t_i}sgn\left( E\left( t_i,\delta\phi \right) \right) \sqrt{\omega \dot{\rho}\left( t_i,\delta\phi \right) }e^{i\omega t_r\left( t_i,\delta\phi \right) },
\label{eq:EH}
\end{multline} where the sum is over all the $t_i(\omega,\delta\phi)$'s that
satisfy $U(t_i,t_r,\delta\phi)+I_p=\hbar\omega$. The sign function $sgn(E)$ is
introduced to ensure the centro-symmetry of the medium, implying that the
radiation from right and left trajectories have opposite polarity. The
ionization density is calculated from $\dot{\rho}(t)=[\rho_0-\rho(t)]w(\lvert
E(t)\rvert)$ where $w(.)$ is the tunneling ionization rate
\cite{Ammosov1986,Scrinzi1999}and $\rho_0$ is the initial gas density. For
simplicity, Eq. \ref {eq:EH} is derived assuming uniform spectral density
of the trajectories, i.e.~$\vert dU/dt_i\vert=$constant, which is a fair
approximation for the plateau  harmonics.
\begin{figure}[htb]
 \centerline{\includegraphics[trim=1cm 1cm 1cm 0.5cm,width=3in]{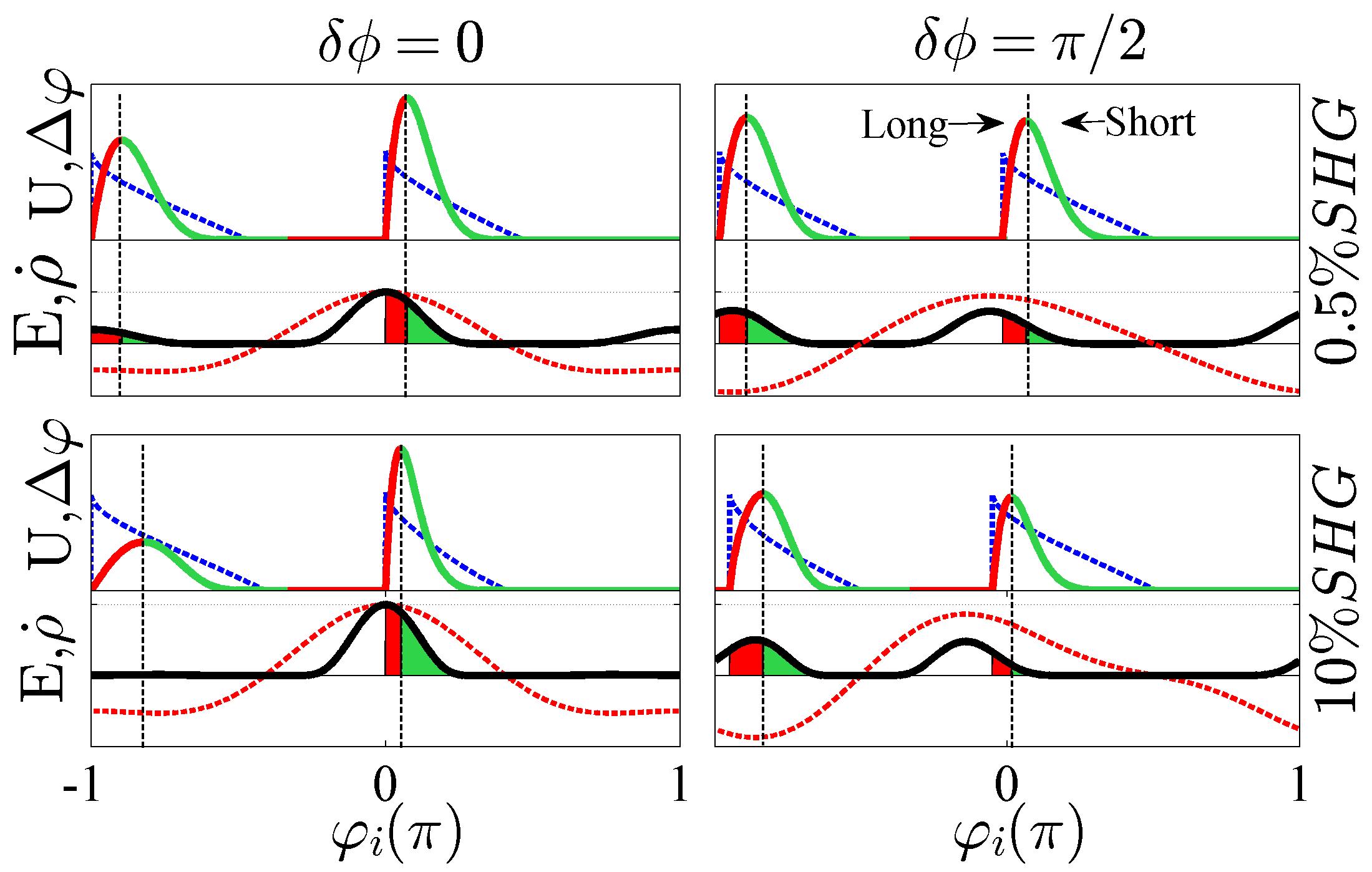}}
 \caption{The electron transit time (phase= $\Delta\varphi/\pi$ ,blue) and return kinetic energy $U$, the electric field $E$ (doted red) and $\dot{\rho}$ (black) versus the ionization time (phase) for two extreme relative phase delays assuming 0.5\% and 10\% SHG for top and bottom rows, respectively. Colored areas represent contributions from long (red) and short (green) trajectories separated by vertical dashed lines.}
 \label{Fig:3step}
 \end{figure}
Calculated transit times
($\Delta\varphi=\omega_0(t_r-t_i)=\varphi_r-\varphi_i$), return energy $U$, and
electric field $E$ are plotted in Fig. \ref{Fig:3step} versus $\varphi_i=\omega_0
t_i$. Similar to the case of one-color excitation, the long and short
trajectories are identified as those corresponding to before and after the peak
of returned energies in each half-cycle, respectively.

\begin{figure*}[ht]
\includegraphics[trim=0cm 1cm 0cm 1cm,width=5.8in]{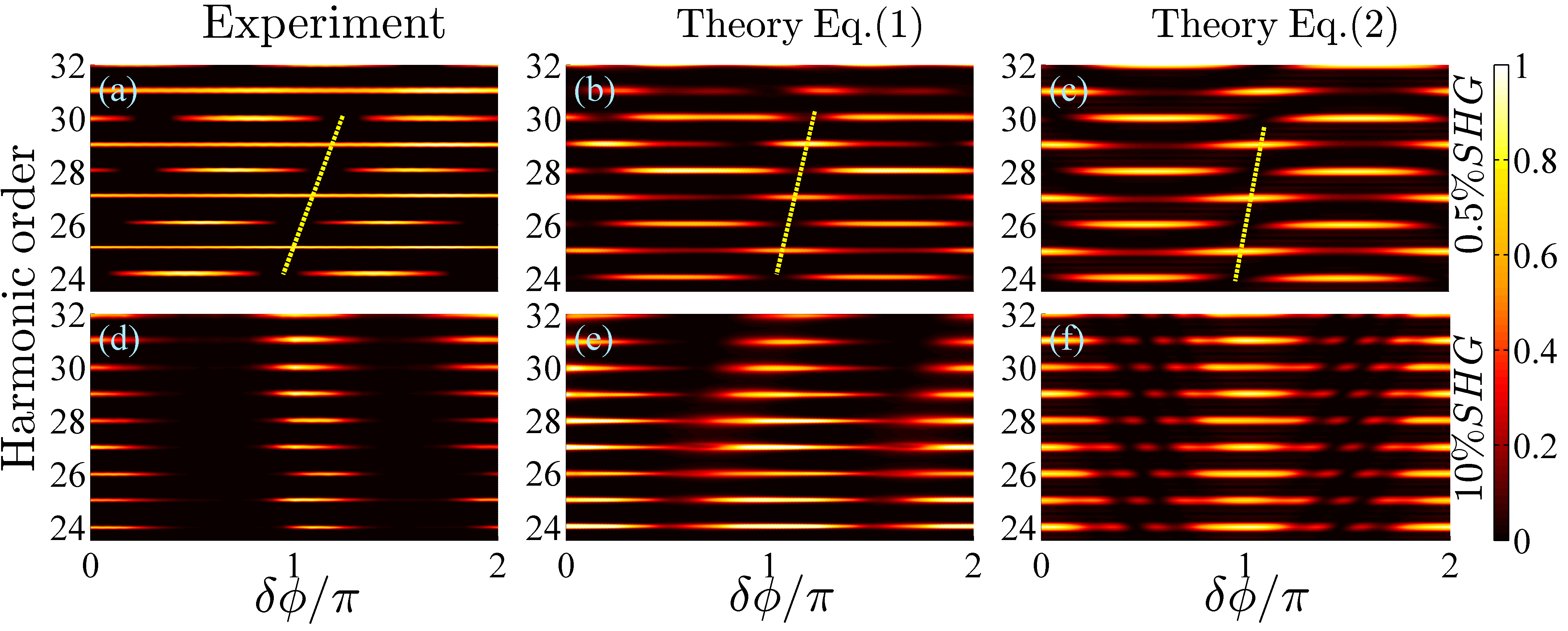}
\caption{Comparison of experimental (a and d) and theoretical [Eq. \ref{eq:EH} (b and e), Eq. \ref{eq:EH2} (c and f)] spectra of XUV emission excited by pulses with weak (0.5\% SHG, top) and strong (10\% SHG, bottom) polar asymmetry.}
\label{Fig:Elowinject}
\end{figure*}

Fig. \ref{Fig:Elowinject} shows the calculated HHG spectra $S\left(\omega ,
\delta\phi\right)=|E_H|^2$ (middle column) using Eq. \ref{eq:EH} in  excellent
qualitative agreement with our measurements (left column) for both weak
($\eta=0.005$) and strong ($\eta=0.1$) polar asymmetry. For typical experimental
conditions in the 2-color excitation scheme, most observations correspond to the
mid-harmonics within the plateau ($1.5I_p<\hbar\omega<1.5U_p$). Given the peak
intensities involved in these experiments ($\approx 5\times10^{14}$  $W/cm^2$), this
corresponds to $\approx$ H16-H26 harmonics of the fundamental at 800 nm. In this
regime, the emission is dominated by the long electron
trajectories as evident from the larger amplitude of $\dot{\rho}$ at the birth
time of these trajectories. The contribution of short trajectories becomes
noticeable only at energies approaching the cut-off (not shown).

The simplicity of our model allows us to identify the underlying physical
mechanisms manifested in the experimental results. In the high polar asymmetry
case (Fig. \ref{Fig:3step} bottom), the extreme nonlinearity in the tunnel
ionization rate suppresses the amplitude ($\dot{\rho}$) in the adjacent half
cycle (for $\delta\phi=m\pi$), thus leading to a full-cycle periodicity of the
HHG signal in the time domain. Fourier analysis shows how this leads
to the formation of both odd and even harmonics. At phase delays corresponding
to $\delta\phi=(2m+1)\pi/2$, the time-domain signal returns to half-cycle
periodicity but with much lower amplitude (Fig. \ref{Fig:3step}). The production
of XUV is thus peaked at $\delta\phi\approx m\pi$, which is also
anti-correlated in $\delta\phi$ dependence with the THz emission process
(Fig. \ref{Fig:XUVTHz}).

More striking is the agreement of this simple model with the results of weak
polar asymmetry (SHG=0.5\%, Fig. \ref{Fig:Elowinject}), where two main features
of our experimental data and others
\cite{Dudovich2006,Doumy2009,Zhang2012,Dahl2011}) are reproduced. We note
that even harmonics appear out-of-phase with the odd harmonics. Another curious
feature is the small shift in the HHG maxima or minima with phase
delay $\delta\phi$, as depicted in Fig. \ref{Fig:Elowinject} with a line of slope
$\approx\pi/50$ or 27 attoseconds per harmonic.  \citet{Dudovich2006} exploited
this shift to deduce the phase of the electron wavepacket and to
control the creation of attosecond pulses. In the case of
weak polar asymmetry, the effect of the ionization rate is not as dominant and
the observed HHG spectral behavior is determined instead by the spectral
interference of emitted radiation from the first and the second half cycles in
each optical period of the fundamental field.
 The injection of the small SH signal breaks the half-cycle periodicity
resulting in the generation of even harmonics in addition to odd harmonics.  The
variation of the spectral phase, $\varphi_q=\omega t_r$, with harmonic order (
$q=\omega/\omega_0$) and the phase delay $\delta\phi$  explains the shift in the
harmonic peaks shown in Fig. \ref{Fig:Elowinject}. The interference between two
adjacent long (and short) trajectories involves the differential spectral phase
$\Delta\varphi_q={(\varphi^{(1)}_r-\varphi^{(2)}_r)q}$ with (1) and (2)
indicating the first and second half-cycles. For a given $\omega$, this quantity
can be evaluated numerically using the classical equations of motion
presented earlier. With no SH injection ($\eta=0$), we have
$\Delta\varphi_q=q\pi$ for both short and long trajectories due to half-cycle
periodicity. Deviation of $\Delta\varphi_q$ from $\pi$ due to SH injection and
its dependence on the phase delay can be evaluated numerically to the first
order. For mid-
harmonics, this variation is fit with the $\Delta\varphi_q^{(L)}(\omega,
\delta\phi)\approx q(\pi+1.33{\sqrt{\eta}}\times\sin{(\delta\phi-0.060q}))$ and
$\Delta\varphi_q^{(S)}(\omega, \delta\phi)\approx
q(\pi+1.8{\sqrt{\eta}}\times\cos{(\delta\phi+0.075q}))$ for long and short
trajectories, respectively. We note that the empirical slope of 0.06 radians
($\approx\pi/50$) per harmonic for $\Delta\varphi_q^{(L)}$is in close agreement
with our measurements and more generally with all the reported experimental
mid-harmonic data. The broad applicability of this finding highlights
the validity of a classical description of the electron trajectories.\\
To further emphasize this point, we simplify the expression in Eq. \ref{eq:EH}
by assuming that $\dot{\rho}$ scales as $\vert E\vert^Q$ and take the SH
injection as a perturbation. A comparison with the static tunnel ionization rate
indicates that $Q\approx$ 7-9 for the laser intensities in these
experiments. Considering the dominant birth times to be near the peaks of each
cycle, two adjacent long trajectories will take relative amplitudes
$A^{\pm}\approx(1\pm \sqrt{\eta}\cos(\delta\phi))^{Q\over2}\approx \exp(\pm
{Q\over2}\sqrt{\eta}\cos(\delta\phi))$. From Eq. \ref{eq:EH}, a square pulse
having $N$ cycles will then produce an HHG spectrum:
\begin{equation}
 S\left(\omega , \delta\phi\right)\propto\left| \sinh\left( {Q\over 2}\sqrt{\eta}\cos\left( \delta\phi\right)-i{\Delta\varphi_q^{(L)}\over 2} \right)  \right|  ^2 C_N(q),
 \label{eq:EH2}
 \end{equation}
where $C_N(q)=\sin^2(N\pi q)/ \sin^2(\pi q)$ is a comb function. A plot of this simple expression assuming $Q$=8 for 0.5\% and 10\% SH injection is shown in right column of Fig. \ref{Fig:Elowinject}, where the main features of the observations are reproduced.  We have assumed $N$=4 cycles for clarity. 
At higher harmonics approaching the cut-off, the amplitude of the short
trajectories becomes comparable to the long trajectories leading to
additional interference and consequently more complex dependence of the HHG
spectra on the phase delay. As indicated by the empirical fits to the
differential phases $\Delta\varphi_q^{(L)}$ and $\Delta\varphi_q^{(S)}$ given
above, the modulation of the HHG spectra by the phase delay for the long and
short trajectories has opposite signs (-0.060 vs +0.075), which causes this
modulation to reverse at higher harmonics. This behavior is in qualitative
agreement with experiments that were explained by the semi-classical
calculations \cite{Dahl2011,Zhang2012}. A more detailed analysis of this regime
is beyond the scope of this letter, and will be addressed in later publications.

The excellent agreement between experiments and the ETS model points to the
robustness of the original 3-step model in its semi-classical approximation of
high-harmonic generation. For the case of strong injection, the ETS model in
combination with the plasma current model verifies the observed anti-correlated
dependence of XUV and THz on the relative phase parameter $\delta\phi$. In the
regime of weak injection, this dependence is more complicated due to an
additional phase shift depending on the harmonic order. In this regime, the
Coulomb potential has been argued to be responsible for deviations in the
correlation of XUV and THz emission from the predictions of the plasma current
model \cite{Zhang2012}. Our results clarify the critical role of classical long
electron trajectories and their differential re-collision times from each
adjacent half-cycle in modulating the XUV spectrum under conditions of both
weak and strong injection. Additionally, the ETS model naturally includes a
unified
description of XUV and THz emission within the framework of classical trajectories, experimentally verified for conditions of strong injection.

In conclusion, we investigated simultaneous emission of THz and XUV radiation
from two-color laser-driven plasmas. By controlling the polar asymmetry of
the synthetic driving field, we observe and quantify correlations between
emission of ultrashort bursts of coherent THz and XUV emission. We presented an
extended 3-step model based on the classical electron trajectories that
provides exceptional agreement with experimentally observed XUV spectra in the
regimes of weak and strong second harmonic injection. Furthermore, the agreement
of the observed correlation features in the THz and XUV emission with our
theory suggests that the two processes have the same physical origin, only with
different electron trajectories.
   
The work at UNM was partially supported by the NSF under Award DMR-1207489, and by the Los Alamos National Laboratory (LDRD) program 20120246ER. DVS acknowledges support by the NSF (Grant No. 1160764) and partial support by the EU FP7 Marie Curie Zukunftskolleg Incoming Fellowship Program, University of Konstanz (Grant No. 291784).  
\bibliographystyle{apsrev4-1}
\bibliography{STX}

\end{document}